\documentclass[10pt, conference, compsocconf]{IEEEtran}

\usepackage{url}
\usepackage{epsfig}
\usepackage{graphics}
\usepackage{epstopdf}
\usepackage{color}
\usepackage{booktabs}
\usepackage{array}
\usepackage{multicol}
\usepackage{placeins}
\usepackage{xspace}
\usepackage[font=small,bf,labelsep=period]{caption}

\newcommand{\bi}{\begin{itemize}}
\newcommand{\ei}{\end{itemize}}
\newcommand{\be}{\begin{enumerate}}
\newcommand{\ee}{\end{enumerate}}
\newcommand{\apenet}{APEnet\xspace}
\newcommand{\apenetp}{APEnet+\xspace}
\newcommand{\PCIe}{PCIe\xspace}
\newcommand{\buflist}{\texttt{BUF\_LIST}\xspace}
\newcommand{\hostvtop}{\texttt{HOST\_V2P}\xspace}
\newcommand{\gpuvtop}{\texttt{GPU\_V2P}\xspace}
\newcommand{\PtoP}{\textit{peer-to-peer}\xspace}
\newcommand{\nios}{\texttt{Nios~II}\xspace}
\newcommand{\ptoptx}{\texttt{GPU\_P2P\_TX}\xspace}
\newcommand{\clusterI}{Cluster~I\xspace}
\newcommand{\clusterII}{Cluster~II\xspace}
\newcommand{\us}{\,$\mu$s\xspace}
\newcommand{\ie}{\textit{i.e.}\xspace}
\newcommand{\eg}{\textit{e.g.}\xspace}
\newcommand{\etc}{\textit{etc.}\xspace}

\ifCLASSINFOpdf
\else
\fi

\begin{document}

\title{GPU peer-to-peer techniques applied to a cluster interconnect}


\author{\IEEEauthorblockN{
Roberto Ammendola\IEEEauthorrefmark{1}, Massimo Bernaschi\IEEEauthorrefmark{2},
Andrea Biagioni\IEEEauthorrefmark{3}, Mauro Bisson\IEEEauthorrefmark{2},
Massimiliano Fatica\IEEEauthorrefmark{4}}
\IEEEauthorblockN{Ottorino Frezza\IEEEauthorrefmark{3}, Francesca Lo Cicero\IEEEauthorrefmark{3}, Alessandro Lonardo\IEEEauthorrefmark{3}, Enrico Mastrostefano\IEEEauthorrefmark{5},
Pier Stanislao Paolucci\IEEEauthorrefmark{3}}
\IEEEauthorblockN{Davide
Rossetti\IEEEauthorrefmark{3},Francesco Simula\IEEEauthorrefmark{3}, Laura
Tosoratto\IEEEauthorrefmark{3} and Piero Vicini\IEEEauthorrefmark{3}}\\

\IEEEauthorblockA{
  \IEEEauthorrefmark{1} Istituto Nazionale di Fisica Nucleare,\\
  Sezione Roma Tor Vergata, Rome, Italy}
  \vspace{0.3cm}
  \IEEEauthorblockA{\IEEEauthorrefmark{2} Istituto Applicazioni Calcolo,\\
    Consiglio Nazionale delle Ricerche, Rome, Italy}
   \vspace{0.3cm}   
 \IEEEauthorblockA{ \IEEEauthorrefmark{3} Istituto Nazionale di Fisica Nucleare,\\
  Sezione Roma, Piazzale Aldo Moro 2, I-00186 Rome, Italy}
    \vspace{0.3cm}
  \IEEEauthorblockA{\IEEEauthorrefmark{4} NVIDIA Corp. CA, USA}
  \vspace{0.3cm}
  \IEEEauthorblockA{\IEEEauthorrefmark{5} Sapienza University,\\
  Department of Computer Science, Via Salaria 113, Rome, Italy}}

\maketitle

\begin{abstract}
Modern GPUs support special protocols to exchange data directly across
the PCI Express bus.
While these protocols could be used to reduce GPU data transmission
times, basically by avoiding staging to host memory, they require
specific hardware features which are not available on current generation network
adapters.
In this paper we describe the architectural modifications required to
implement \PtoP access to NVIDIA Fermi- and
\mbox{Kepler-class} GPUs on an \mbox{FPGA-based} cluster interconnect.

Besides, the current software implementation, which integrates this
feature by minimally extending the RDMA programming model, is
discussed, as well as some issues raised while employing it in a
higher level API like MPI.

Finally, the current limits of the technique are studied by analyzing
the performance improvements on low-level benchmarks and on two
GPU-accelerated applications, showing when and how they seem to
benefit from the GPU \PtoP method.
\end{abstract}

\section{Introduction}

Thanks to their high computational power and high memory bandwidth, as
well as to the availability of a complete programming environment,
Graphics Processing Units (GPUs) have quickly risen in popularity and
acceptance in the HPC world.
The challenge is now demonstrating that they are suitable for
capability computing, \ie they are still effective when scaling on
\mbox{multi-node} \mbox{multi-GPU} systems.
\mbox{Multi-GPU} is needed to decrease \mbox{time-to-solution} (strong
scaling), to match the requirements of modern demanding applications, \eg
by overcoming the GPU memory size limitations (bigger simulation
volumes, finer grained meshes, \etc) or even to enable new types of
computation which otherwise would not be possible.

In many cases, applications show poor scaling properties because, on
increasing the number of computing elements, the computation locally
performed by processing nodes shrinks faster than the amount of
communication.
To fix it, a well-established optimization technique is \emph{overlapping} computation
and communication.
Additionally, due to \emph{staging} of GPU data in host buffers prior and/or
after the communication phase, network exchange of large buffers
typically needs a proper coding (pipelining) to obtain good bandwidth.
These two techniques are inter-related: \emph{staging}, when not
properly implemented, can hurt \emph{overlapping} due to unexpected
synchronizations of GPU kernels; on the other hand, \emph{staging}
data, while computing is underway, is essential to obtain the
\emph{overlap}.

The use of the \PtoP exchange of data among multiple GPUs in a single
box, instead of using a simple intra-node message-passing approach,
is reported to provide a 50\% performance gain on capability problems, as
for example recently discussed in the literature~\cite{MGPUFORPHYS}.
In principle, the same \PtoP technology, applied to an
interconnection network, can also be employed to enable remote
transfers of GPU buffers without staging to host memory.
Anyway, a few capabilities are required to do so efficiently, at least
on NVIDIA Fermi-class GPUs, which are beyond those available in
current commercial cluster interconnects (InfiniBand, 10G).
Those capabilities, if cost effective, will probably appear in next
generation silicon devices with the typical delay due to VLSI design
and production cycles (18-24 months).

In the meantime, it is entirely possible to experiment with GPU
\PtoP networking by using reconfigurable components,
like the Altera Statix IV in the \apenetp card~\cite{APENETP2P}, which
offers both \mbox{high-performance} specialized transmission logic
blocks and \mbox{high-capacity} on-chip memory banks.

In this paper we report on our experiences in adding GPU
\PtoP capabilities to the \apenetp network
interconnection, presenting some early performance results.

The paper is organized as follows: in section~\ref{sec:rel} we cite
some related works; section~\ref{sec:bgnd} briefly introduces the GPU
\PtoP technology and the \apenetp network adapter, while
section~\ref{sec:p2ponnet} describes the implementation issues related
to how the technology has been introduced in \apenetp;
section~\ref{sec:bmarks} contains the preliminary results obtained on
\apenetp, in particular on synthetic benchmarks and on two
\mbox{multi-GPU} applications; the last section contains conclusions
and some final remarks.

\section{Related works}
\label{sec:rel}

Previous attempts at integrating GPUs with network interconnect
middlewares, although showing interesting results, all fall in the
\mbox{software-only} category.

While early attempts at integrating \mbox{GPU-awareness} with MPI date
back to 2009~\cite{cudampi2009}, two of the most widely used MPI
implementations, OpenMPI~\cite{OPENMPI} and MVAPICH2~\cite{MVAPICH2},
have recently started to offer the possibility of specifying GPU
memory pointers in MPI \mbox{point-to-point} functions, suppressing
the chore of explicitly coding the data transfers between GPU and CPU
memories.
This feature represents an essential part of the research efforts
aimed towards the definition of a general mechanism for direct
communication among GPUs.
%
Despite that, both OpenMPI and MVAPICH2 rely on a software approach
which eases programming and that can increase communication
performance for \mbox{mid-to-large-size} messages, thanks to pipelining
implemented at the MPI library level.
On the other hand, this approach can even hurt performance~\cite{BENCH} for
\mbox{medium-size} messages, due to them not using independent CUDA
STREAMs, thereby introducing an implicit synchronization that ruins
the \mbox{computation-communication} overlap in
applications.

On a related note, MPI-ACC \cite{MPIACC} experimented with alternative
approaches at integrating GPUs with MPI, even discovering unexpected
\mbox{slow-downs} and bugs in previous CUDA releases.

\section{Background}
\label{sec:bgnd}

Fermi is the first NVIDIA GPU architecture which externally exposes a
proprietary \mbox{HW-based} protocol to exchange data among GPUs
directly across the PCI Express bus (\PCIe), a technique which is
generically referred to as \PtoP, and publicly advertised under the
GPUDirect Peer-to-peer moniker.

The NVIDIA \PtoP protocol comprises a number of hardware resources
(registers, mailboxes) implemented on the GPU and set of rules to use
them. 
This protocol basically allows one GPU to read and write the memory of
another GPU, provided that they are on a compliant platform (suitable
\PCIe bus topology, bug-free chipsets).
At the API level, it is used by the \texttt{cudaMemcpyPeer()} and
\texttt{cudaMemcpyPeerAsync()} APIs, or simply by
\texttt{cudaMemcpy()} on selected platforms (UVA mode), when memory
pointers refers to device memory residing on two different GPUs.
This protocol can be used by a third-party device to gain direct
access to the GPU memory, provided that it is able to consistently
mimic the correct hardware behaviour and that the right bindings to the
NVIDIA kernel driver are established at the software level.
Since CUDA 4.1, the SW support to GPU \PtoP has been
shipped as an internal preview technology, subject to NDA.

Beyond the \PtoP protocol, there is an additional access method for
third-party devices, the so called BAR1.
This methods is alternative to the \PtoP, and CUDA 5.0 has officially
introduced a public API to support it on Kepler-based Tesla and Quadro
GPUs.
With BAR1, it is possible to expose, or \textit{map}, \ie to make it
available, a region of device memory on the second \PCIe memory-mapped
address space of the GPU, from which it can be read or written with
standard \PCIe memory operations.
Due to platform constraints (32-bits BIOS), this address space is
limited to a few hundreds of megabytes, so it is a scarce resource.
Additionally, mapping a GPU memory buffer is an expensive operation,
which require a full reconfiguration of the GPU.
As it is shown below, the BAR1 reading bandwidth on Fermi is quite
limited, suggesting that the Fermi architecture was not optimized for
this access method.
That is why we have never used the BAR1 method on Fermi.

\apenetp has been supporting \PtoP on Fermi since the end of 2011, and
since recently also BAR1 and \PtoP on Kepler.

\subsection{GPU \PtoP technology}

In CUDA, the \PtoP/BAR1 support for third-party device is split
into a user- and a \mbox{kernel-space} part.
The \mbox{user-space} function \texttt{cuPointerGetAttribute()} with
the \texttt{CU\_POINTER\_ATTRIBUTE\_P2P\_TOKENS} parameter is used to
retrieve special handles from pointers to GPU memory buffers.
Those handles are then used in \mbox{kernel-space} to properly map the
GPU buffers, \ie one page descriptor for each 64~KB page, comprising
the physical page address plus additional low-level protocol tokens
which are used to physically read and write GPU memory.

%
Technically, \PtoP writing of GPU memory is only slightly more
difficult than host memory writing, the only difference being the
managing of a sliding window to access different pages.
%
GPU memory reading is instead more complex because it is designed
around a \mbox{two-way} protocol between the initiator and the target
devices. This design is justified by the need to work around bugs in
\PCIe chip-sets, related to a traffic pattern among devices which
is still quite uncommon at least on the x86 platform.
The ability to use the \PtoP protocol among GPUs, and its performance,
is constrained by the \PCIe topology; performance is excellent when
two GPUs share the same \PCIe \mbox{root-complex}, \eg they are
directly connected to a \PCIe switch or to the same hub.
Otherwise, when GPUs are linked to different bus branches,
performance may suffers or malfunctionings can arise.
This can be an issue on \mbox{multi-socket} Sandy Bridge Xeon
platforms, where \PCIe slots might be connected to different
processors, therefore requiring GPU \PtoP traffic to
cross the \mbox{inter-socket} QPI channel(s).

\subsection{\apenetp}

\begin{figure}[h!]
  \centering
  \includegraphics[width=.8\columnwidth]{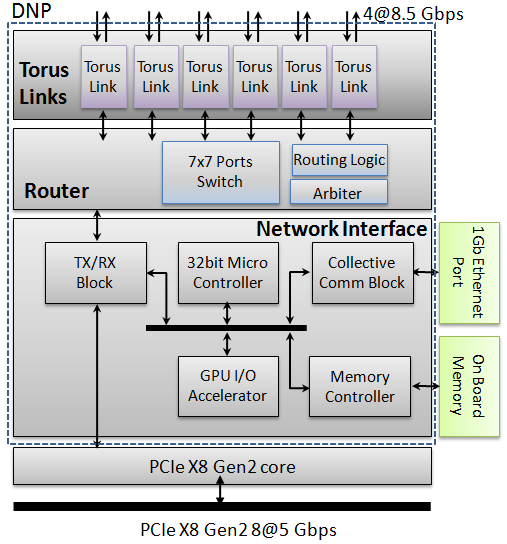}
  \caption{Overview of \apenetp. The DNP is the core of the
    architecture --- composed by the Torus Links, Router and Network
    Interface macro-blocks --- implemented on the FPGA. The system
    interfaces to the host CPU through the PCIe bus.}
  \label{fig:internals}
\end{figure}

\apenet is a 3D Torus interconnection technology originally proposed,
in its first version, back in 2004~\cite{Ammendola2005826} and which
is now being developed in its second generation version, called
\apenetp~\cite{Ammendola:2010uq}.
It has a direct network design which combines the two traditional
components: the Network Interface (NI) and the Router (RTR).
The Router implements a \mbox{dimension-ordered} static routing
algorithm and directly controls an \mbox{8-ports} switch, with 6 ports
connecting the external torus link blocks ($X^+$, $X^-$, $Y^+$, $Y^-$,
$Z^+$, $Z^-$) and 2 local packet injection/extraction ports.
%
The \apenetp Network Interface comprises the PCIe X8 Gen2 link to the
host system, for a maximum data transfer rate of 4+4~GB/s, the packet
injection logic (TX) with a 32~KB transmission buffer, and the RX RDMA
logic which converts the virtual memory address of the destination
buffer into a scatter list of physical memory addresses.
The core architecture is depicted on Fig~\ref{fig:internals}.
%
%
The receiving (RX) data path manages buffer validation (the \buflist)
and \mbox{virtual-to-physical} address translation (the \hostvtop
map);
these tasks are currently partly implemented in software running on a
\mbox{micro-controller} (\nios), which is synthesized onto the Stratix
IV FPGA.
On the transmit (TX) data path, equivalent tasks are carried on by the
kernel device driver, which implements the message fragmentation and
pushes transaction descriptors with validated and translated physical
memory addresses.

The \apenetp architecture is designed around a simple Remote Direct
Memory Access (RDMA) programming model.
The model has been extended with the ability to read and write the GPU
private memory --- global device memory in CUDA wording --- directly
over the PCIe bus, by exploiting the NVIDIA GPUDirect
\PtoP (P2P) HW protocol.

\section{Implementing GPU \PtoP technology on \apenetp}
\label{sec:p2ponnet}

\apenetp is relatively easy to extend thanks to the presence of the
reconfigurable hardware component (an Altera FPGA) which, among other
resources like transceiver blocks, memory banks, \etc, provides a
32~bit \mbox{micro-controller} (\nios) that can run up to 200~MHz and
is easily programmable in C.

Introducing GPU \PtoP in \apenetp has been relatively easy for the
receive data path, in that the 64~KB page GPU windowing access has been
implemented as a variation of the 4~KB page host memory writing flow.
Either the relevant data structures have been extended (the \buflist)
to accept both host and GPU buffers, or new ones have been added (the
new \gpuvtop map, one per GPU).

For both read and write, physical GPU memory addresses are needed to
generate the transactions on the \PCIe link, so a proper GPU
\mbox{virtual-to-physical} address translation module, \gpuvtop, has
been implemented on \apenetp, very similar to but not exactly the same
as the host one.
For each GPU card on the bus, a \mbox{4-level} GPU\_V2P page table is
maintained, which resolves virtual addresses to GPU page descriptors.

Currently, the processing time of an incoming GPU data packet is of
the order of 3~\us (1.2~GB/s for 4~KB packets) and it is equally
dominated by the two main tasks running on the \nios: the \buflist
traversal (which linearly scales with the number of registered buffers) and the
address translation (which has constant traversal time thanks to the
\mbox{4-level} page table).

\begin{figure}[htbp]
\hspace{20pt}
\centering
\includegraphics[width=.4\textwidth]{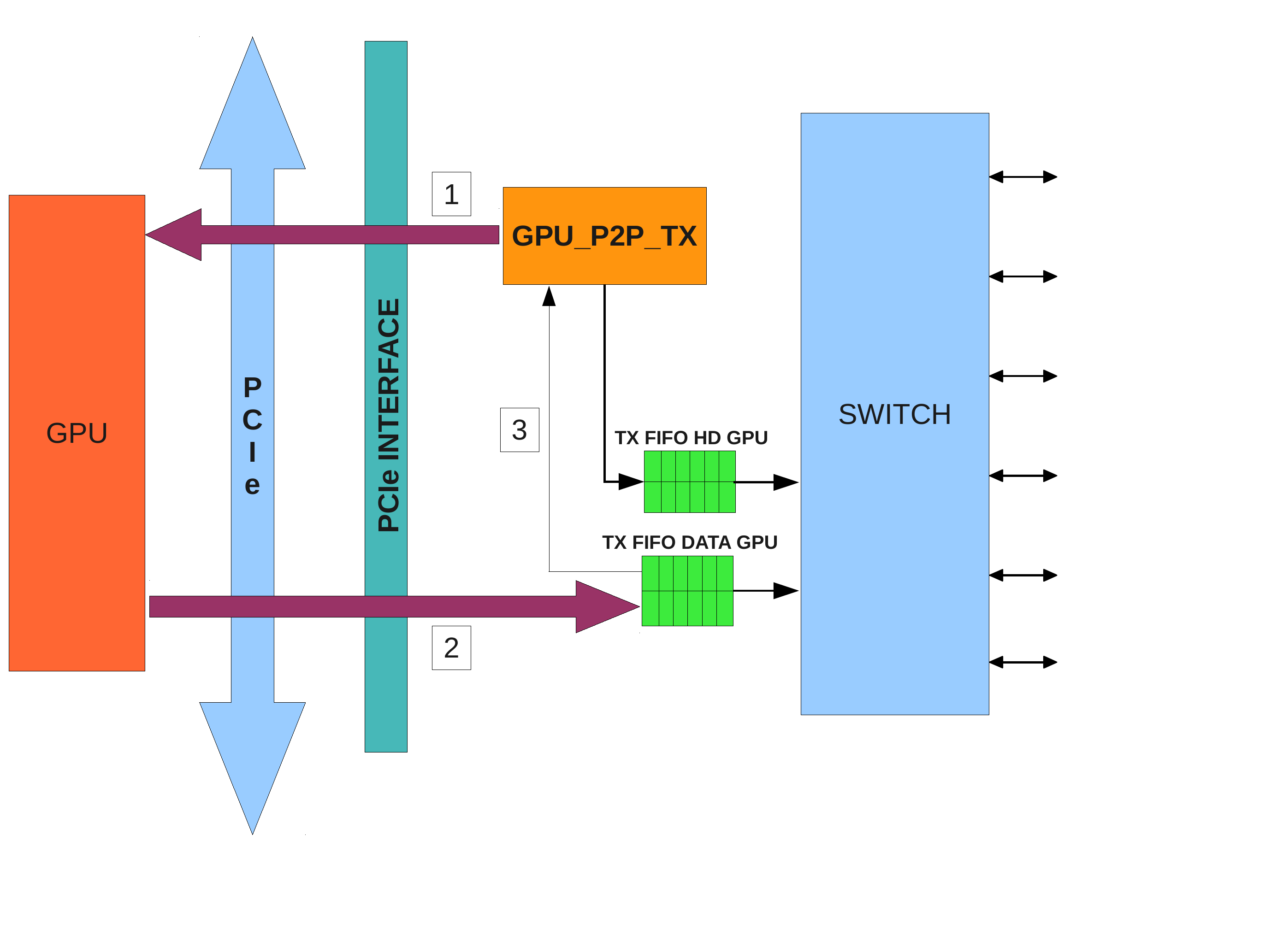}
\caption{A part of the \ptoptx hardware blocks in action: read
  requests (1) generation logic, in-flight GPU data packets (2),
  flow-control feed-back (3).}
\label{fig:gpu_p2p}
\end{figure}
Good performance in GPU memory reading, which is used during
transmission of GPU buffers, has instead been by far the most difficult
task to achieve, requiring two major redesigns
of the related SW/HW blocks.
Unlike host buffer transmission, which is completely handled by the
kernel driver, GPU data transmission is delegated to the \apenetp.
This is so not only to \mbox{off-load} the host, which would be a
minor requirement, but mainly for the architectural need of
maintaining the correct data flow among the different actors involved:
the \mbox{multiple-outstanding} read request queue of the GPU (arrow 1
in fig.~\ref{fig:gpu_p2p}), the data flow (arrow 2) insisting on the
\apenetp data transmission buffers and the outgoing channel buffers.
This flow needs proper management to avoid buffer overflowing but at
the same time it has to be carefully tuned to obtain enough
performances.
With the first generation GPU memory reading and control flow logic
(\ptoptx) (V1), which was able to process a single packet request of
up to 4KB, the peak GPU reading bandwidth~\cite{APENETP2P} was
throttled to 600~MB/s.
The reasons for the poor performances were: the slow rate of read
requests emitted by the \ptoptx block towards the GPU and the long
latency of the GPU in responding with data, which is quite
understandable as the GPU memory subsystem is optimized for throughput
rather than for latency.
Besides, the \ptoptx was impacting the RX processing path due to the
high computation load on the \nios \mbox{micro-controller}.

The second generation \ptoptx implements two key improvements: an
hardware acceleration block which generates the read requests towards
the GPU with a steady rate of one every 80~ns; a pre-fetch logic which
attempts to hide the response latency of the GPU.
Additionally, thanks to the acceleration blocks, the \nios
\mbox{micro-controller} can allot a larger \mbox{time-slice} to the
receive data path (RX processing).
By using a 32~KB prefetch window, which is related to the size of the
transmission buffers (TX FIFO), the \ptoptx was able to reach the
current peak of 1.5~GB/s for the GPU reading bandwidth.

In the last generation \ptoptx (V3), the \textit{new flow-control}
block is able to \mbox{pre-fetch} an unlimited amount of data so as to
keep the GPU read request queue full, while at the same time
\mbox{back-reacting} to \mbox{almost-full} conditions (arrow 3 in
fig.~\ref{fig:gpu_p2p}) of the different \mbox{on-board} temporary
buffers (TX Data FIFO, TX Header FIFO, \PtoP REQUEST FIFO, \etc).

Fig. \ref{fig:gpu_tx_prefetch_bw} and \ref{fig:gpu_prefetch_bw}
show the effect of the different \ptoptx implementations onto the GPU
reading bandwidth and on the whole loop-back bandwidth.

\subsection{Changes to RDMA API for GPUs} 

The \apenetp APIs have been extended to handle transmission and
reception of GPU buffers, in a way that makes extensive use of the
Uniform Virtual Address (UVA) capability of CUDA.
With UVA --- available on most \mbox{64-bits} platforms and OS's, ---
GPU buffers are assigned unique \mbox{64-bits} addresses, and they can
be distinguished from plain host memory pointers by using the
\texttt{cuPointerGetAttribute()} call, which also returns other
important buffer properties like the GPU index and the CUDA context.

The \apenetp buffer pinning and registration API now accepts GPU
buffers, which are mapped on-the-fly if not already present in an
internal cache.
Buffer mapping consists in retrieving the \PtoP informations, then
passing them down to the kernel driver and from there to the \nios
\mbox{micro-controller}, in the \buflist and \gpuvtop data structures.
After registration, a buffer --- either a host or GPU, uniquely
identified by its (UVA) \mbox{64-bit} virtual address and process ID
--- can be the target of a PUT operation coming from another node.
Network packets carry the \mbox{64-bit} destination virtual memory
address in the header, so when they land onto the destination card,
the \buflist is used to distinguish GPU from host buffers.

On the transmitting node, the source memory buffer type is chosen at
compilation time by passing a flag to the PUT API.
This is useful to avoid a call to \texttt{cuPointerGetAttribute()},
which is possibly expensive~\cite{MPIACC}, at least on early CUDA 4
releases.
When a GPU buffer is transmitted, the buffer mapping is automatically
done, if necessary, and a simplified descriptor list containing only
GPU virtual addresses is generated by the kernel driver and passed to
the \nios \mbox{micro-controller}.
As in the RX processing phase, the \nios is once again in charge of
the \mbox{virtual-to-physical} translation of source memory page
addresses, in addition to driving the GPU \PtoP protocol
together with the HW acceleration blocks.

\section{Benchmarking}
\label{sec:bmarks}

In this section we report the results of GPU peer-to-peer enabled
benchmarks and applications on \apenetp.

The \apenetp test platform (\clusterI in the following) is made of
eight \mbox{dual-socket} Xeon Westmere nodes, arranged in a $4\times2$
torus topology, each one equipped with a single GPU (all Fermi 2050
but one 2070) and a Mellanox ConnectX-2 board, plugged in a \PCIe X4
slot (due to motherboard constraints) and connected to a Mellanox
MTS3600 switch.

Infiniband results were collected on a second 12-nodes Xeon Westmere
cluster (\clusterII), each node equipped with two Fermi 2075 GPUs
(Tesla S2075) and a Mellanox ConnectX-2 board, plugged in a \PCIe X8
slot and connected to a Mellanox IS5030 switch.

ECC is off on both clusters.
MVAPICH2 1.9a2 and OSU Micro Benchmarks v3.6 were used for all MPI IB
tests.

\subsection{\PCIe bus analysis}

%
%
%
%
\begin{figure}[htbp]
\centering
\includegraphics[width=.4\textwidth]{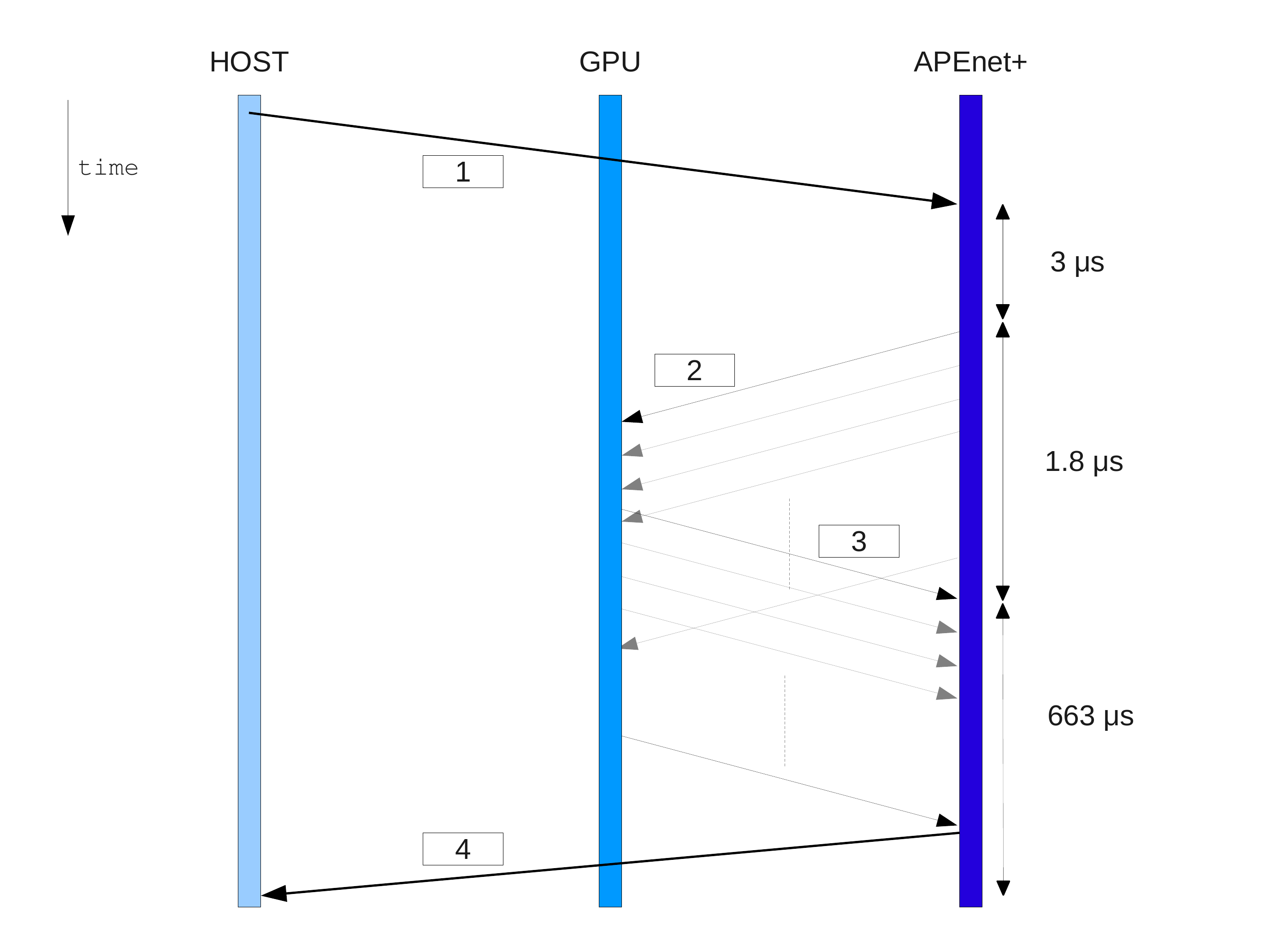}
\caption{Sketch of the \PCIe timings related to \PtoP transactions.
  \ptoptx v2 with 32~KB pre-fetch window.  Test is successive
  transmission of a single 4~MB GPU buffer. Bus Analyzer with a \PCIe
  X8 Gen2 active interposer. SuperMicro 4U server with PLX \PCIe
  switch. }
\label{fig:pcie_timings}
\end{figure}
The successive revisions of the \apenetp \PtoP support has been guided
by a low-level analysis of the performance on the \PCIe bus, through
the use of a bus analyzer.

In fig.~\ref{fig:pcie_timings}, we report the timings of the most
important \PCIe bus transactions, as seen by an active interposer
sitting between the \apenetp card and the motherboard slot.
The most interesting informations are: the \apenetp \ptoptx (v2, in
this case) implementation has an overhead which is a substantial part
of those 3~\us in the initial delay (transaction 1 to 2). That overhead
partially overlaps with previous transmissions, so it is paid in
full either at the beginning of a long communication phase (with minor
effects) or on short-message round-trip tests (with visible effects on
the network latency).

The head reading latency of GPU is 1.8~\us (transaction 2 to 3), then
it sustains a 1536~MB/s data throughput towards \apenetp transmission
buffers (transaction 3 to 4 is 663~\us for a single 1~MB message, 53\%
link utilization).
The read requests generated towards the GPU by the \ptoptx hardware
accelerator are regularly emitted once every 76~\us, \ie 96~MB/s of
protocol traffic and 13\% link utilization.

As of the \PtoP write bandwidth, judging from \PCIe bus traces, the
GPU has no problem sustaining the PCIe X8 Gen2 traffic, even though
\apenetp is currently not able to use all the available bandwidth due
to limitations of its RX packet processing (more on this below).

\subsection{\mbox{Single-node} benchmarks}

To give an idea of the performance and limitations of the current
implementation, in table~\ref{tab:lowlevel} we collected the memory
read performance, as measured by the \apenetp device, for buffers
located on either host or GPU memory.
\begin{small}
\begin{table}[htbp]
\centering
\setlength\extrarowheight{2pt}
\setlength{\tabcolsep}{1pt}
\begin{tabular}{|m{2cm}|l|l|l|}
\hline
\textbf{Test}  & \textbf{Bandwidth} & \textbf{GPU/method} & \nios active tasks\\
\hline
Host mem read  & 2.4~GB/s &             & none \\
GPU mem read   & 1.5~GB/s & Fermi/P2P   & \ptoptx \\
GPU mem read   & 150~MB/s & Fermi/BAR1  & \ptoptx \\
GPU mem read   & 1.6~GB/s & Kepler/P2P  & \ptoptx \\
GPU mem read   & 1.6~GB/s & Kepler/BAR1 & \ptoptx \\
\hline
GPU-to-GPU loop-back   & 1.1~GB/s  & Fermi/P2P  & \ptoptx + RX\\
\hline
Host-to-Host loop-back & 1.2~GB/s  &            & RX \\
\hline
\end{tabular}
\caption{\apenetp \mbox{low-level} bandwidths, as measured with a
  \mbox{single-board} loop-back test. The memory read figures have
  been obtained by flushing the packets while traversing \apenetp
  internal switch logic.  BAR1 results taken on an ideal platform,
  \apenetp and GPU linked by a PLX \PCIe switch.  Kepler results are
  for a pre-release K20 (GK110), with ECC enabled. Fermi results are
  with ECC off. GPU and \apenetp linked by a PLX \PCIe switch.}
\label{tab:lowlevel}
\end{table}
\end{small}
As discussed in the previous section, the complexity of the GPU
\PtoP read protocol and the limitations of our implementation
set a limit of 1.5~GB/s to the Fermi GPU memory read bandwidth, which is
roughly half that obtained for host memory read (2.4~GB/s). 
For reference, the GPU-to-host reading bandwidth, as obtained by
\texttt{cudaMemcpy}, which uses the GPU DMA engines, peaks at about
5.5~GB/s on the same platform.
We also report very early results on Kepler GPUs, for both K10 and
K20, which show a 10\% increase in the available \PtoP reading
bandwidth with respect to Fermi in P2P mode, and a more impressive
factor 10 using the BAR1 approach (150~MB/s on Fermi vs. 1.6~GB/s on
K20).
Tests on Kepler GPUs have been run on pre-release cards, so the
reported performance is subject to change.

We underline that this is the reading bandwidth as measured from
\apenetp through the GPU \PtoP protocol, neither the internal device
bandwidth, which is instead available to kernels running on the GPU,
nor the GPU DMA engine bandwidth, \eg \texttt{cudaMemcpy()}.

%

The last two lines of table~\ref{tab:lowlevel} show that, when the
packet RX processing is taken into account by doing a loop-back test,
the peak bandwidth decreases from 2.4~GB/s to 1.2~GB/s in the
host-to-host case, and from 1.5~GB/s to 1.1~GB/s in the GPU-to-GPU case,
\ie an additional 10\% price to pay in the latter case.
The last column in the table shows that the \nios
\mbox{micro-controller} is the main performance bottleneck.
%
We are currently working on adding more hardware blocks to accelerate
the RX task.

\begin{figure}[!htb]
\centering
\includegraphics[width=.45\textwidth,clip=true,trim=0 50pt 0 50pt]{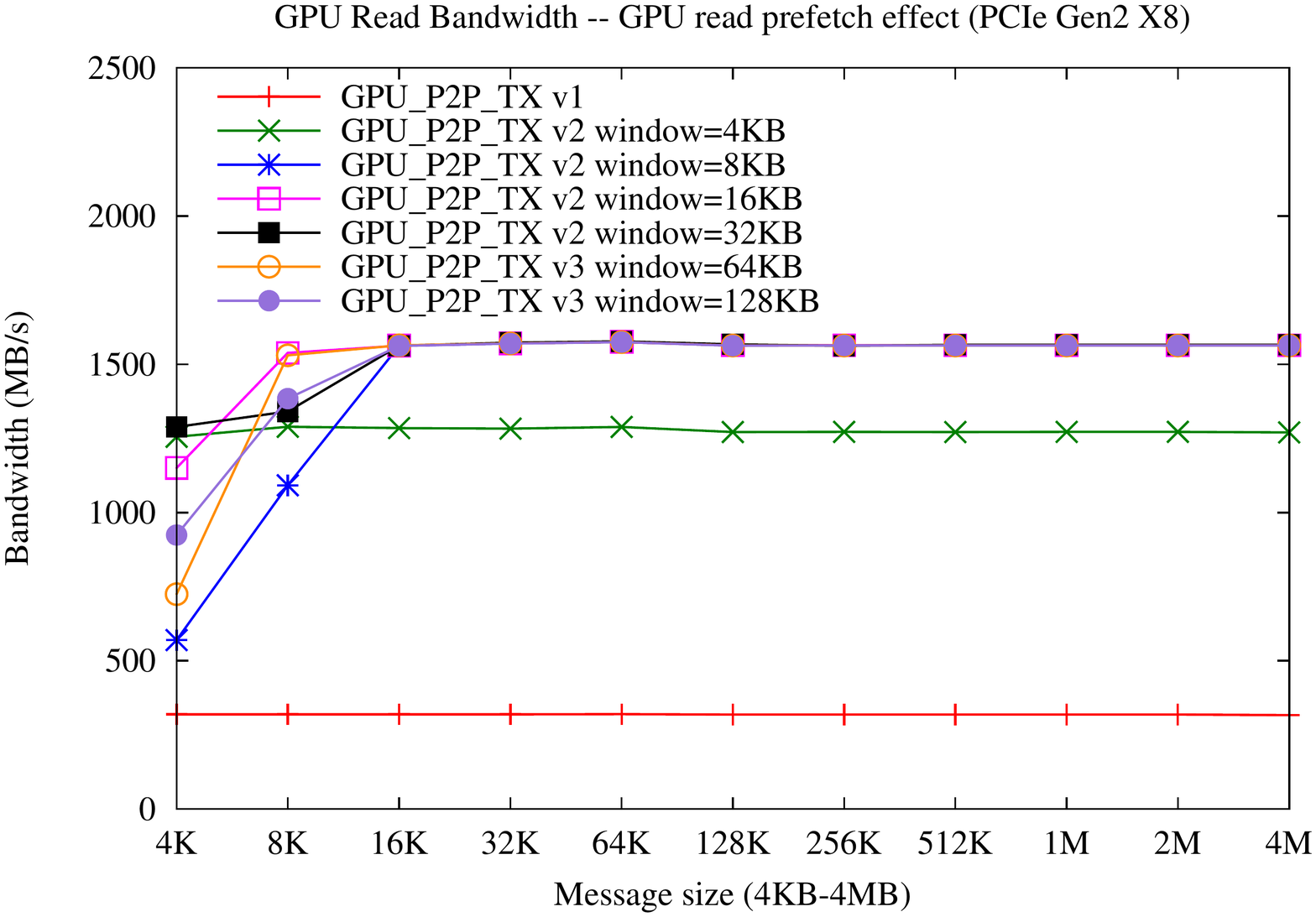}
\caption{Single-node GPU memory reading bandwidth, showing the
  performance at varying message size, obtained by flushing TX
  injection FIFOs. Different curves corresponds to the three \ptoptx
  implementations and to different pre-fetch window sizes, where
  appropriate. Plots are not smooth for small message sizes 
  due to software related issues under queue-full conditions. }
\label{fig:gpu_tx_prefetch_bw}
\includegraphics[width=.45\textwidth,clip=true,trim=0 50pt 0 50pt]{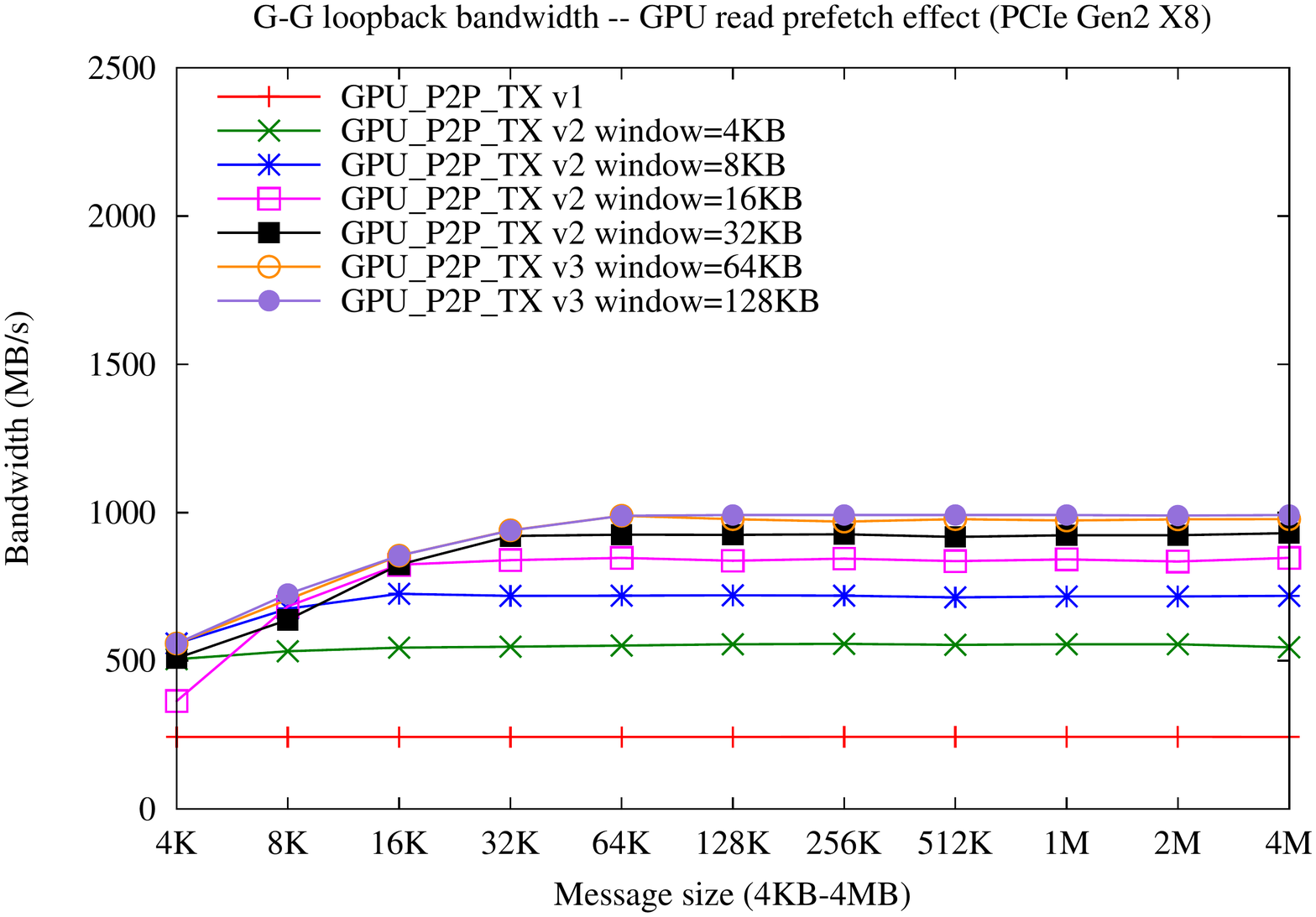}
\caption{Single-node GPU memory loop-back bandwidth, at varying
  pre-fetch threshold size. Different curves are as in the the
  previous plot. The full loop-back send-and-receive bandwidth is
  plotted, which is limited by the \nios \mbox{micro-controller}
  processing capabilities. }
\label{fig:gpu_prefetch_bw}
\end{figure}
%
%
%
%
The values reported in table~\ref{tab:lowlevel} are obtained as the
peak values in a loop-back performance test, coded against the
\apenetp RDMA API.
The test allocates a singe receive buffer (host or GPU), then it
enters a tight loop, enqueuing as many RDMA PUT as possible as to keep
the transmission queue constantly full.
%
%
Fig.~\ref{fig:gpu_tx_prefetch_bw} is a plot of GPU reading bandwidth
at varying message sizes, estimated by using the test above and by
flushing TX injection FIFOs, effectively simulating a zero-latency
infinitely fast switch.
The original \ptoptx v1 implementation (no pre-fetching and
software-only implementation on \nios) shows its limits. 
\ptoptx v2 (HW acceleration of read requests and limited pre-fetching)
shows a 20\% improvement while increasing the pre-fetch window size
from 4KB to 8KB.
%
Unlimited pre-fetching and more sophisticated flow-control in \ptoptx
v3 partially shows its potential only in the full loop-back plot of
Fig.~\ref{fig:gpu_prefetch_bw}
Here the \nios handles both the \ptoptx and the RX tasks, so therefore
any processing time spared thanks to a more sophisticated GPU TX
flow-control logic reflects to an higher bandwidth.
This also suggests that the \apenetp bi-directional bandwidth, which
is not reported here, will reflect a similar behaviour.

\subsection{\mbox{Two-nodes} benchmarks}
%
%
\begin{figure}[!htbp]
\centering
\includegraphics[width=.45\textwidth,clip=true,trim=0 50pt 0 50pt]{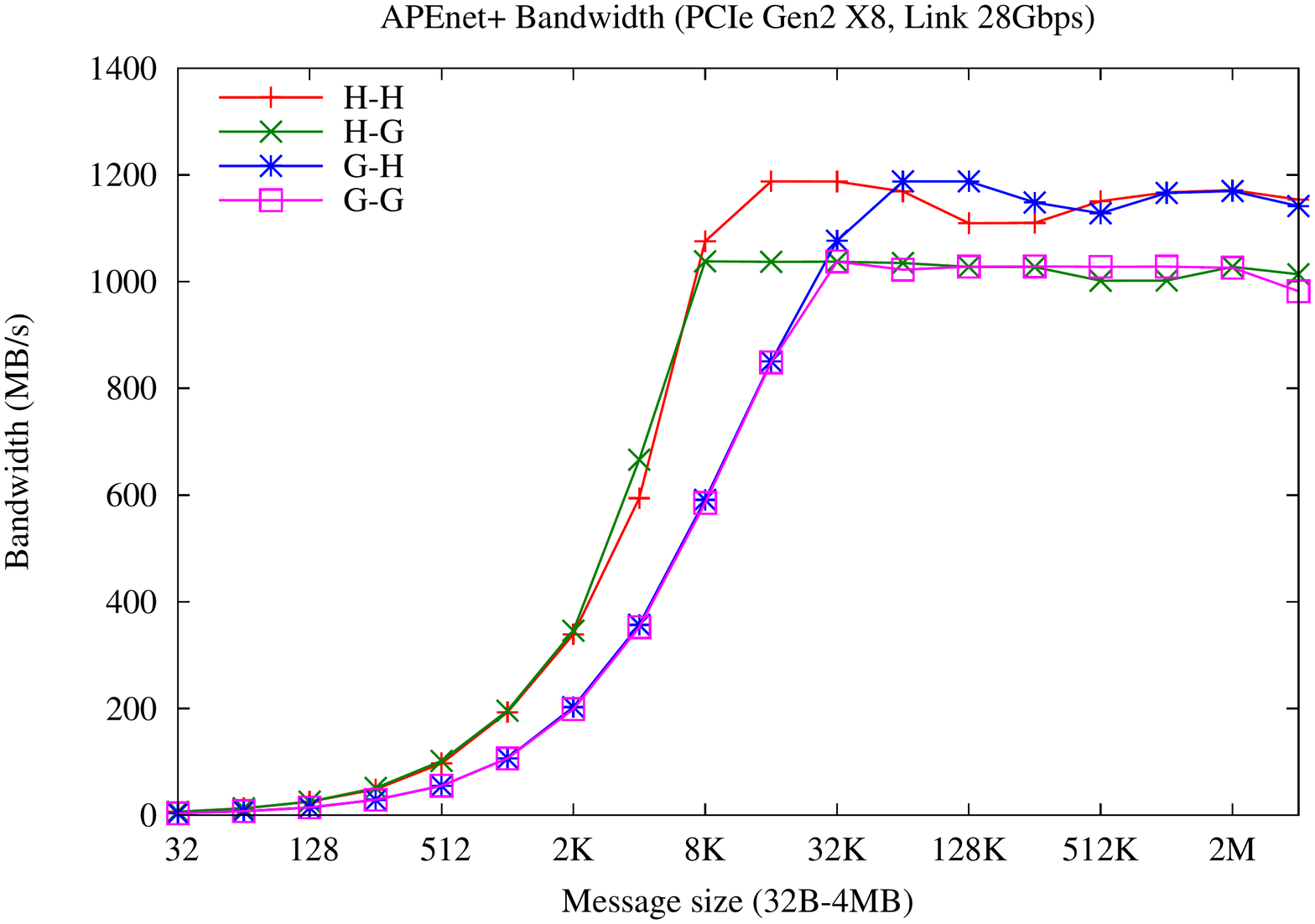}
\caption{\mbox{Two-nodes} \mbox{uni-directional} bandwidth test, for
  different combinations of both the source and the destination buffer
  types. When source is in GPU memory, the overhead is visible; at
  8KB, the bandwidth is almost half that in the host memory case. The
  bandwidth cap is related to the limited processing capabilities of
  the \nios \mbox{micro-controller}. }
\label{fig:apenet_bw}
\includegraphics[width=.45\textwidth,clip=true,trim=0 50pt 0 50pt]{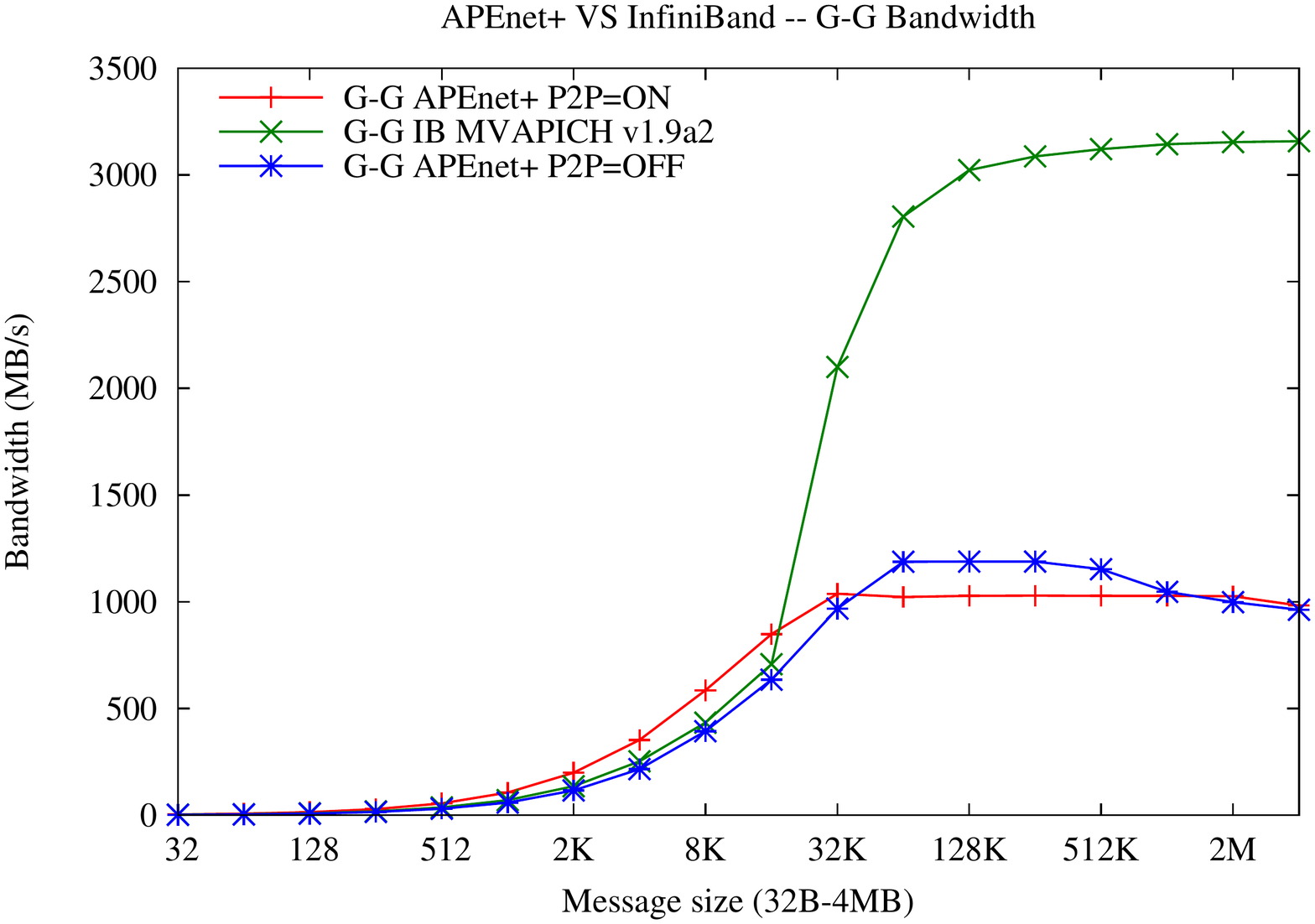}
\caption{\mbox{Two-nodes} \mbox{uni-directional} bandwidth test,
  GPU-to-GPU. P2P=OFF case corresponds to the use of staging in host
  memory. MVAPICH2 result on OSU MPI bandwidth test is for reference.}
\label{fig:apenet_vs_ib_bw}
\end{figure}
%
%
As shown above, reading bandwidth from GPU memory and RX processing
are the two key limiting factors of the current \apenetp
implementation.
Therefore, it can be expected that they influence the communication
bandwidth between two nodes in different ways, depending of the type
of the buffers used.
To measure the effect of those factors independently, we run a two
node bandwidth test on \apenetp, in principle similar to the MPI
OSU~\cite{OMB-GPU} uni-directional bandwidth test, although this one
is coded in terms of the \apenet RDMA APIs.

The plot in Fig.~\ref{fig:apenet_bw} shows the bandwidth of \apenetp
for the four different possible combinations of source and destination
buffer types:
for source buffers located in host memory, the best performance of
1.2~GB/s is reached, with a 10\% penalty paid when receive buffers are
on the GPU, probably related to the additional actions involved, \ie
switching GPU \PtoP window before writing to it.
For GPU source buffers, the GPU \PtoP reading bandwidth is the
limiting factor, so the curves are less steep and only for larger
buffer sizes, \ie beyond 32~KB, the plateau is reached.
Clearly, the asymptotic bandwidth is limited by the RX processing, but
the overall performance is affected by the transmission of GPU
buffers.
Interestingly, the Host-to-GPU performance seems to be a very good
compromise bandwidth-wise, \eg for 8~KB message size the bandwidth is
twice that of the GPU-to-GPU case.
Of course this plot is good for analyzing the quality of the \apenetp
implementation, but it says nothing about which method is the best for
exchanging data between GPU buffers, \ie in which ranges GPU \PtoP is
better than staging on host memory.
To this end, Fig.~\ref{fig:apenet_vs_ib_bw} is a plot of the
GPU-to-GPU communication bandwidth, with three different methods:
\apenetp using GPU \PtoP; \apenetp with staging of GPU data to host
memory; OSU bandwidth test, using MVAPICH2 over Infiniband, which uses
a pipelining protocol above a certain threshold, used for reference.
The GPU \PtoP technique is definitively effective for small buffer
sizes, \ie up to 32~KB; after that limit, staging seems a better
approach.
%
%
\begin{figure}[!htbp]
\centering
\includegraphics[width=.45\textwidth,clip=true,trim=0 50pt 0 50pt]{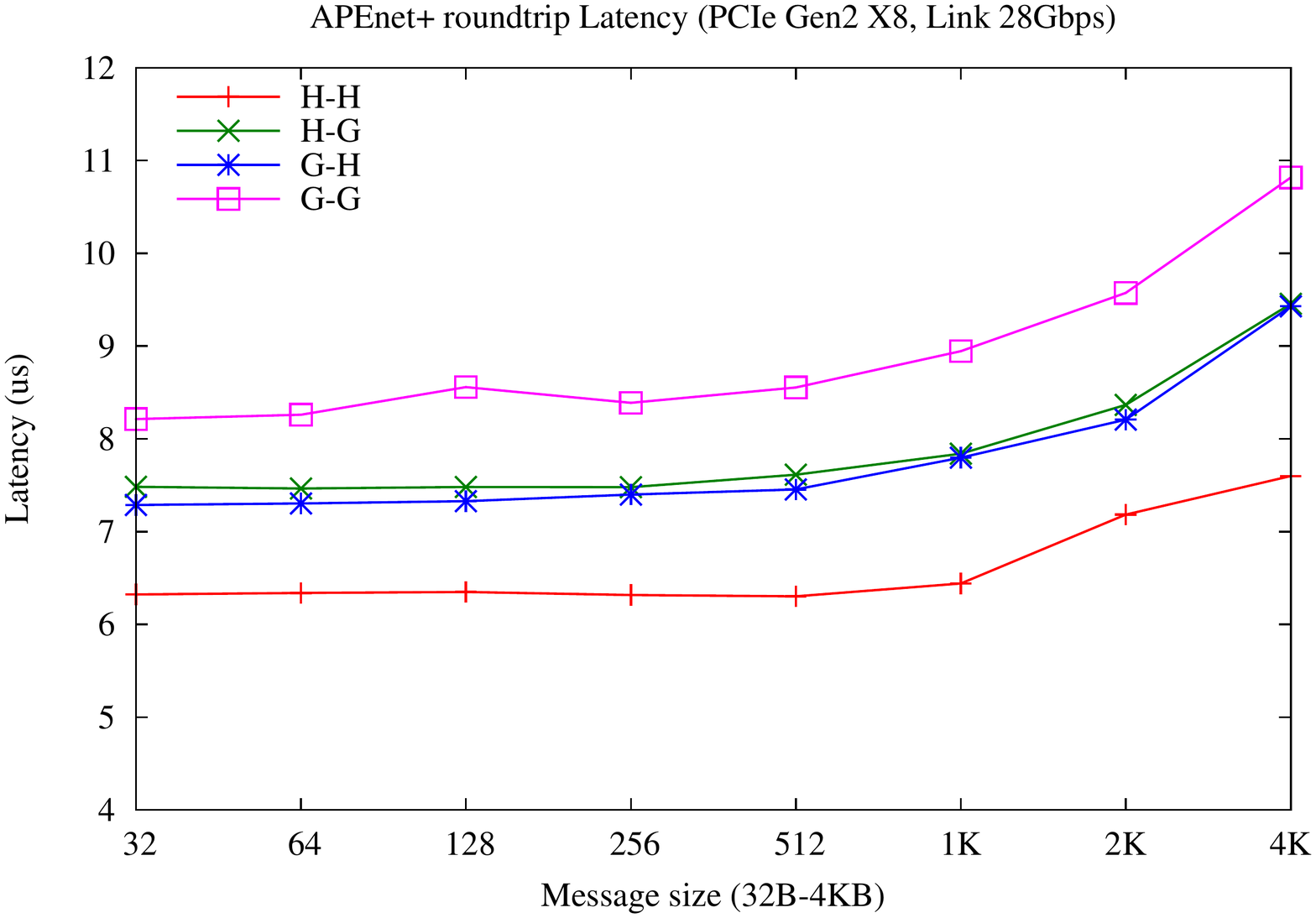}
\caption{\apenetp latency, estimated as half the round-trip
  latency. Different combinations of both the source and the
  destination buffer types.}
\label{fig:apenet_lat}
\includegraphics[width=.45\textwidth,clip=true,trim=0 50pt 0 50pt]{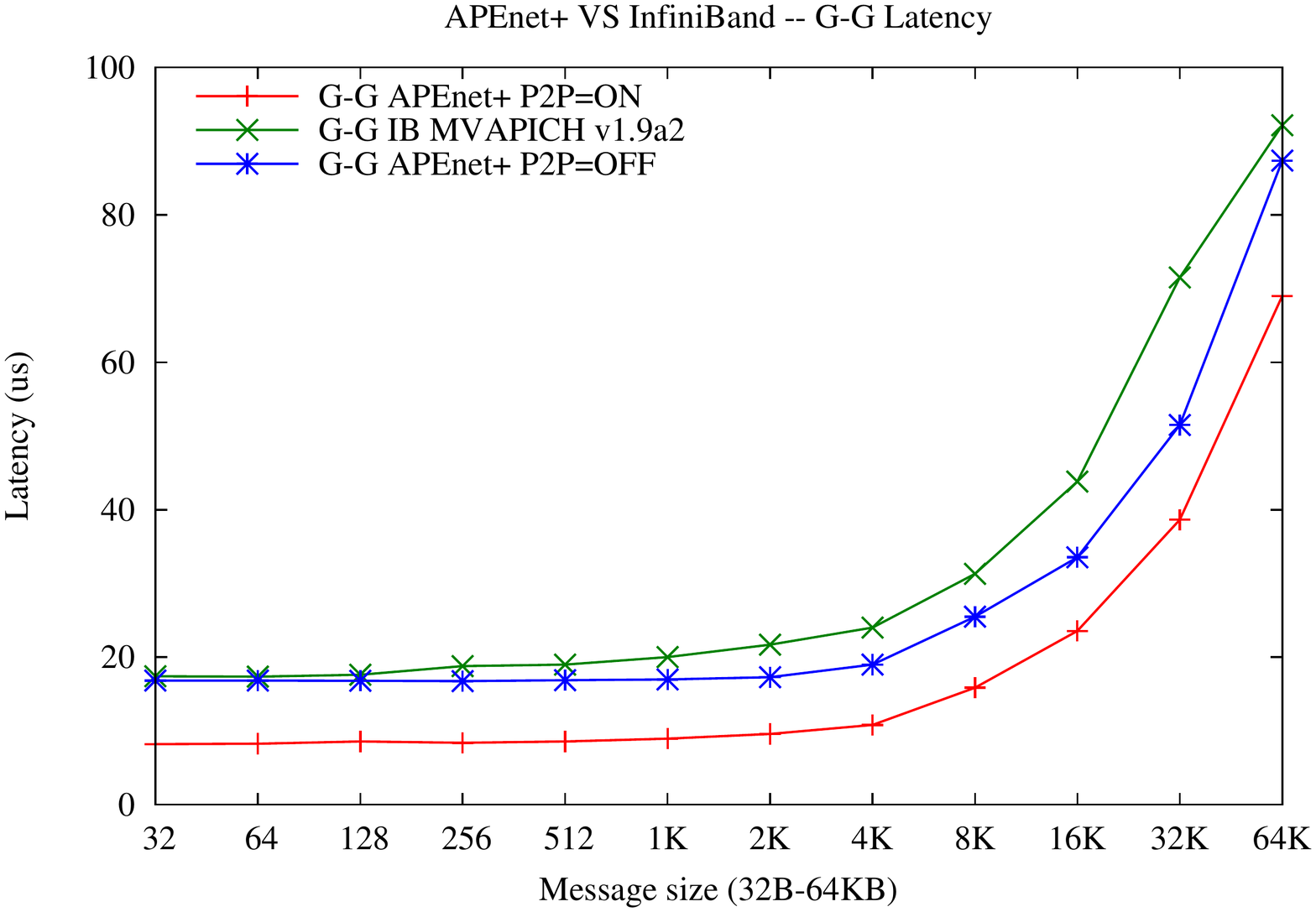}
\caption{\apenetp latency. GPU-to-GPU case. \PtoP has 50\% less
  latency than staging. The MVAPICH2 plot is the GPU OSU latency test
  on Infiniband.}
\label{fig:apenet_vs_ib_lat}
\includegraphics[width=.45\textwidth,clip=true,trim=0 50pt 0 50pt]{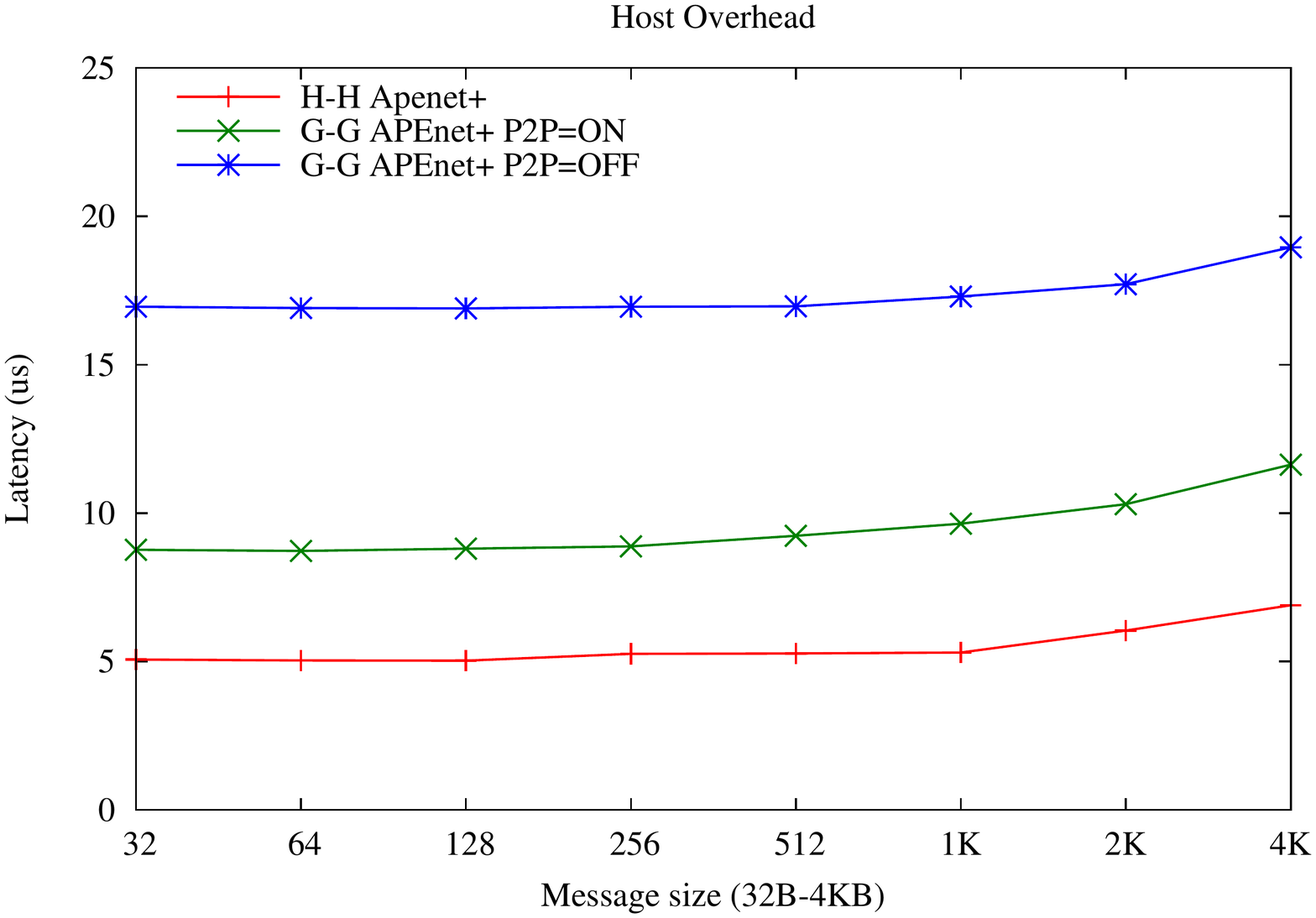}
\caption{\apenetp \textit{host overhead}, estimated via bandwidth test.}
\label{fig:host_overhead}
\end{figure}
%
%
%
Fig.~\ref{fig:apenet_vs_ib_lat} is more useful to explore the
behaviour of GPU \PtoP on small buffer size.
Here the latency, estimated as half the round-trip time in a ping-pong
test, shows a clear advantage of the \PtoP implementation with respect
to staging (P2P=OFF in the figure), even on a very low-latency network
as Infiniband.
Indeed, the \apenetp \PtoP latency is 8.2~\us, while for \apenetp with
staging and MVAPICH2/IB it is respectively 16.8~\us and 17.4~\us. 
In the latter case, most of the additional latency comes from the
overhead of the two CUDA memory copy (\texttt{cudaMemcpy}) calls
necessary to move GPU data between temporary transmission buffers.
By subtracting the \apenetp H-H latency (6.3~\us in
Fig.~\ref{fig:apenet_lat}) from the \apenetp latency with staging
(16.8~\us), the single \texttt{cudaMemcpy} overhead can be estimated
around 10~\us, which was confirmed by doing simple CUDA tests on the
same hosts.

The run times of the bandwidth test, for short message size, are plot
in Fig.~\ref{fig:host_overhead}.
In the LogP model~\cite{LogP}, this is the host overhead, \ie the
fraction of the whole message send-to-receive time which does not
overlap with subsequent transmissions.
Of those 5~\us in the host-to-host case, at least a fraction can be
accounted to the RX processing time (3~\us estimated by cycle counters
on the \nios firmware).
The additional 3~\us in the GPU-to-GPU (P2P=ON) case should be quite
related to \PtoP protocol as implemented by \apenetp, \eg the
3+1.8~\us \ptoptx overhead in Fig.~\ref{fig:pcie_timings}.
%
When staging is used instead (P2P=OFF), out of the additional 12~\us
(17-5~\us of the host-to-host case), at least 10~\us are due to
the \texttt{cudaMemcpy} device-to-host, which is fully synchronous
with respect to the host, therefore it does not overlap.

In conclusion, the GPU \PtoP, as implemented in \apenetp, shows a
bandwidth advantage for message sizes up to 32~KB.
Beyond that threshold, at least on \apenetp it is convenient to give
up on \PtoP by switching to the staging approach.
Eventually that could have been expected, as architecturally GPU \PtoP
cannot provide any additional bandwidth, which is really constrained
by the underling PCI-express link widths (X8 Gen2 for both \apenetp and
Infiniband) and bus topology.

\subsection{Over-relaxation in 3D Heisenberg Spin Glass}

%
%
\begin{figure}[!htbp]
\hspace{-20pt}
\centering
\includegraphics[width=.45\textwidth,clip=true,trim=0 50pt 0 50pt]{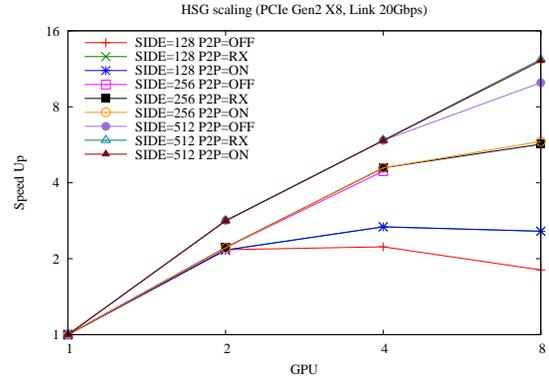}
\caption{HSG: strong scaling. Speedup on \clusterI at varying number
  of computing nodes, for different lattice sizes $L$. For each $L$ we
  show three variants, relative to the use of P2P (off, RX only, RX
  and TX). At $L=512$ a super linear speedup is observed. }
\label{fig:hsg_scaling}
\end{figure}
In this section we show an early evaluation of GPU \PtoP
networking on a \mbox{multi-GPU} simulation code for the Heisenberg
Spin Glass model~\cite{HSG,MGPUFORPHYS}.
Thanks to its regular communication pattern, we consider it a good
model application for typical \mbox{lattice-based} \mbox{multi-GPU}
simulations.
The GPU part of the code is highly optimized; it uses \mbox{even-odd}
parallel update trick;
the 3D domain is decomposed among the computing nodes along a single
dimension, and the \mbox{communication-computation} overlap method is
used: first compute the local lattice boundary, then exchange it with
the remote nodes, while computing the bulk.
The computation consists of multiple \mbox{over-relaxation} steps
applied to the whole spin lattice of size $L^3$:

\begin{table}[hbtp]
\centering
\setlength\extrarowheight{2pt}
\begin{tabular}{|l|ccc|}
\hline
\hline
NP & $T_{tot}$ & $T_{bnd}+T_{net}$ & $T_{net}$ \\
\hline
1  & 921       & 11          & n.a.   \\
2  & 416       & 108         &  97    \\
4  & 202       & 119         & 113    \\
8  & 148       & 148         & 141    \\
%
%
\hline
\hline
\end{tabular}
\caption{HSG: on \clusterI, single-spin update time in picoseconds,
  strong scaling on \apenetp, $L=256$, GPU \PtoP
  networking enabled for both RX and TX.}
\label{tab:hsg_strong_scaling}
\end{table}
In table~\ref{tab:hsg_strong_scaling} we collected the strong-scaling
results on \apenetp for the lattice size $256^3$; times are for
single-spin update in picoseconds, the lower the better.
As expected for the domain decomposition on a single dimension, the
boundary calculation and network communication part is constant while
the bulk computation part shrinks; we expect a good scaling up to
eight nodes, when the two contribution become equal.
%
%
%
%
%
\begin{small}
 \begin{table}[htbp]
 \centering
 \setlength\extrarowheight{2pt}
 \setlength{\tabcolsep}{1pt}
 \begin{tabular}{|l|m{1.5cm}|m{1cm}|m{1.1cm}|m{1.2cm}|m{1.2cm}|}
 \hline
 Time  & \multicolumn{3}{|c|} \clusterI  & \clusterII \newline OMPI \newline (\PCIe~X8) & \clusterI \newline OMPI \newline (\PCIe~X4)\\
 \cline{2-4}
 &P2P=ON  &    P2P=RX & P2P=OFF&&\\  
 \hline
 $T_{tot}$        & 416    & 416  & 416  & 416  & 416  \\
 $T_{bnd}+T_{net}$ & 108    & 97   & 122  & 108  & 108  \\
 $T_{net}$        & 97     & 91   & 114  & 101  & 101  \\
 \hline
 \end{tabular}
 \caption{HSG: on \clusterI, break-down of \apenetp results on two
   nodes; $L=256$; times are picoseconds per single-spin
   update. P2P=off means staging for both TX and RX. P2P=RX is using
   staging for TX and \PtoP for RX only. OpenMPI over Infiniband results
   as reference.}
 \label{tab:hsg_breakdown}
 \end{table}
\end{small}
To better understand the contributions from computation and
communication to the overall performance, in
table~\ref{tab:hsg_breakdown} we collected the results on a
\mbox{two-nodes} \apenetp system:
single spin update times are reported for 3 different
combinations of use of GPU \PtoP.
%
$T_{tot}$ is the total compute time; $T_{bnd}$ refers to the boundary
computation, carried out on an independent CUDA stream respect to the
bulk computation; $T_{net}$ is the communication time alone.
%
%
%
Interestingly, for $L=256$ and two nodes, the bulk computation is long
enough to completely hide the boundary calculation and the
communication.
%
In this case, where for each computation the bulk of the communication
consists of 6 outgoing and 6 incoming 128~KB messages, using the \PtoP
for both TX and RX (P2P=ON) or only for RX (P2P=RX) respectively give
a 14\% and 20\% advantage with respect to the staging approach
(P2P=OFF).
%
%
%
More generally, for $L=128$, the spin lattice is small and comfortably
fit in a single GPU, so it only scales up to 2 nodes.
%
As seen above, $L=256$ scales well up to 4 nodes.
At $L=512$, it scales well up to eight nodes, and a super-linear
speedup is observed, due to strong GPU cache effects.
Indeed, in this case, the spin lattice is so big that it only fits in
a single 2070 6~GB GPU (2050 has only 3~GB), though in this case with
low efficiency (1471~ps for $L=512$ vs 921~ps for $L=256$).
In this case, the P2P=RX case is 28\% better than the staging case.

Acknowledging the fact that the results are subject to change on
different platforms, for different choices of middleware and
application parameters, we can anyway state that GPU \PtoP on \apenetp
is giving a 20-10\% advantage over staging.
This advantage could increase for a multi-dimensional
domain-decomposition, where the size of the exchanged messages shrinks
in the strong scaling, thanks to more regularly shaped 3D sub-domains.

\subsection{\mbox{GPU-accelerated} BFS traversal on distributed systems}

Recent works \cite{Merrill:2011}\cite{Hong:CPU} have shown that, by
using a Level Synchronous BFS, a \mbox{single-GPU} implementation can
exceed in performance \mbox{high-end} \mbox{multi-core} CPU systems.
To overcome the GPU memory limitation, two of the authors (M.B., E.M)
proposed~\cite{ourJPDC} a \mbox{multi-GPU} code that is able to
explore very large graphs (up to 8 billion edges) by using a cluster
of GPU connected by InfiniBand.
We recently modified such code to use GPU
peer-to-peer~\cite{BFSonAPEnet} and, although \apenetp is still in a
development and testing stage, the results, albeit preliminary, show
an advantage with respect to InfiniBand.

%
Executing a BFS traversal on a distributed memory platform, like a
cluster of GPUs, is interesting for several reasons. 
It generates irregular computation and communication patterns: the
typical traffic among nodes can be hardly predicted and, depending on
the graph partitioning, easily shows an all-to-all pattern. 
The messages size varies as well during the different stages of the
traversal, so that the performance of the networking compartment is
exercised in different regions of the bandwidth
plot~\ref{fig:apenet_vs_ib_bw}.
When the size of the graph grows, it is necessary to use more GPUs due
to the limited amount of memory available on a single GPU. 
However, the computation carried out on each GPU increases slowly
whereas the communication increases with the size of the graph and the
number of GPUs, so the improvement to the communication efficiency that
a direct GPU to GPU data exchange may provide is of special
importance.

%

%
\begin{small}
\begin{table}[btp]
\centering
\setlength\extrarowheight{2pt}
\begin{tabular}{|c|c|c|}
\hline
\hline
NP      &  \apenetp \clusterI &  OMPI/IB \clusterII    \\
\hline
1       &  $ 6.7 \times 10^7$  &  $6.2 \times 10^7$      \\
2       &  $ 9.8 \times 10^7$  &  $7.8 \times 10^7$      \\
4       &  $ 1.3 \times 10^8$  &  $8.2 \times 10^7$      \\
8       &  $ 1.7 \times 10^8$  &  $2.0 \times 10^8$      \\
\hline
\hline
\end{tabular}
\caption{BFS: Traversed Edges Per Second, Strong Scaling, number of
  graph vertices $|V|=2^{20}$. \apenetp P2P=ON. Infiniband results are
  for reference.}
\label{tab:ss}
\end{table}
\end{small}
According to the specs of the \texttt{graph500} benchmark
\cite{graph500}, we use, as a performance metrics, the number of
Traversed Edges Per Second (TEPS), so that higher numbers correspond
to better performances.
Our preliminary results, for P2P=ON case only, are summarized in table
\ref{tab:ss}.
Table \ref{tab:ss} shows the strong scaling (the size of the graph is
fixed) obtained for a graph having $2^{20}$ vertices and compares the
results on the two clusters.
\apenetp performs better than InfiniBand up to four nodes/GPUs; after
that point we speculate that the current implementation of the \apenetp 3D Torus network suffers
on this kind of all-to-all traffic.
This is as of now the topic of further investigations.
\begin{figure}[htbp]
  \centering
  \includegraphics[width=.6\columnwidth]{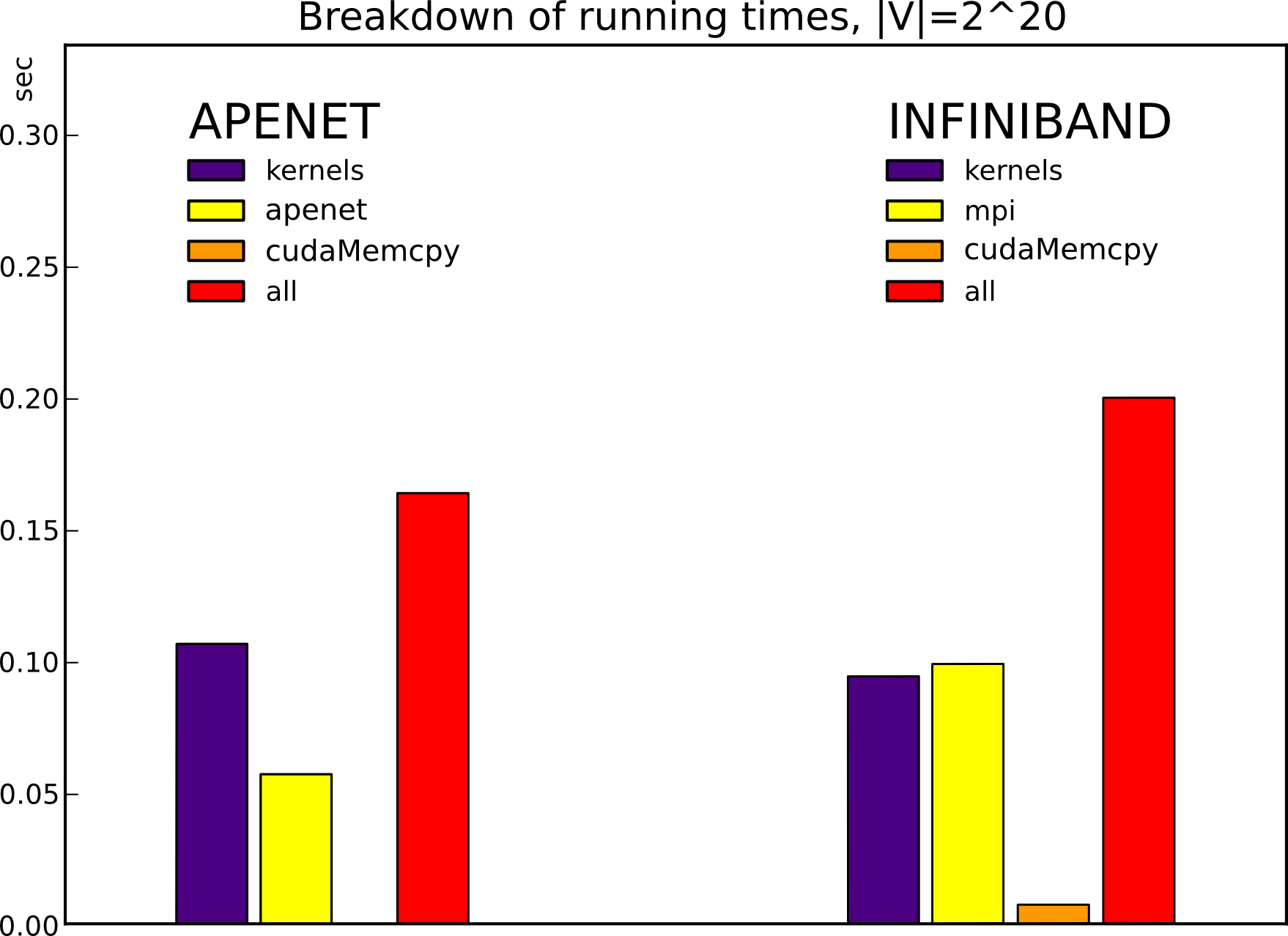}
  \caption{Break-down of the execution time on one out of four tasks
    for both \apenet and InfiniBand.}
  \label{fig:breakdown}
\end{figure}
Although all CUDA kernels and the rest of the code are identical in
the \mbox{MPI-InfiniBand} and the \apenetp version of the code, we
checked that the difference in performances is actually due to the
communication part.
%
%
In the four-nodes case, for this particular graph traversal, the
communication time is 50\% lower in the \apenetp case
(figure~\ref{fig:breakdown}).
%
%

\section{Conclusions\label{sec:conc}}

As it often happens, it is not easy to draw definitive conclusions
on the effectiveness of GPU \PtoP, as it is strongly influenced by the
maturity and efficiency of the particular \apenetp implementation.

Anyway, we can state that the GPU \PtoP write protocol is quite
effective; it has a small overhead and need minor modifications with
respect to writing host memory.

On the other end, the \PtoP reading protocol is complicated for
third-party devices, though minor technical modifications could
improve it a lot; in some sense, it seems too close to the internal
fine-grained memory architecture of the GPU.
Moreover, the reading bandwidth limit around 1.5~GB/s (on Fermi) seems
architectural, verified both at the \PCIe transaction level and by the
scaling with the pre-fetch window.
On the other hand, the GPU \PtoP is by design more resilient to host
platform idiosyncrasies, like \PCIe bus topology and chip-sets bugs.

On Kepler, the BAR1 technique seems more promising. 
In many ways it supports the normal \PCIe protocol for memory-mapped
address spaces, both for reading and writing, so it requires minimal
changes at the hardware level.
The drawback is in platform support, as the \PCIe split-transaction
protocol among devices is known to be deadlock-prone, or at least
sub-performing, on some \PCIe architectures.
Judging from our early experience, the BAR1 reading bandwidth could
be positively affected by the proximity of the GPU and the third-party
device, \eg both being linked to a PLX \PCIe switch.

As of the GPU \PtoP implementation on \apenetp, it seems to be effective
especially in latency-sensitive situations. 
As synthetic benchmarks have shown, \apenetp is able to outperform IB
for \mbox{small-to-medium} message sizes when using GPU \PtoP.
%
%
The advantage provided to the applications by this technique depends
on several factors related to their communication pattern,
i.e. message sizes, destination nodes, source and receive buffer types
(host or GPU) etc., which in turn depend on simulation parameters like
volume size, number of GPUs per node and number of cluster nodes. It
depends also on the possibility of overlapping computation and
communication.
Anyway \PtoP on \apenetp should provide a boost in strong scaling
situations, where the communication pattern is usually dominated by
small-size messages.
Unfortunately, we are currently limited to an 8-nodes test environment;
This is going to change in the next few months, when we will be able
to scale up to 16/24 nodes.


\section*{Acknowledgments}
The \apenetp development is partially supported by the EU Framework
Programme 7 project EURETILE under grant number 247846.

We thanks Timothy Murray, Joel Scherpelz and Piotr Jaroszynski for
supporting our work, and Christopher Lamb for reviewing the paper.

\end{document}